\documentclass[conference]{IEEEtran}
\usepackage{graphicx}
\usepackage{caption}
\usepackage{refstyle}
\usepackage{amsmath}
\usepackage{algorithm}
\usepackage{algorithmic}
\usepackage{epstopdf}

\ifCLASSINFOpdf
\else
\fi
\begin{document}
\title{\huge Successive Optimization Tomlinson-Harashima Precoding Strategies for Physical-Layer Security in Wireless Networks \vspace{-0.5em}}

\author{\IEEEauthorblockN{Xiaotao Lu}
\IEEEauthorblockA{Department of Electronis\\
University of York\\
Email: xtl503@york.ac.uk \vspace{-1.15em}} \and
\IEEEauthorblockN{Rodrigo de Lamare}
\IEEEauthorblockA{CETUC/PUC-Rio\\
Department of Electronis, University of York\\
Email: rodrigo.delamare@york.ac.uk \vspace{-1.75em}} \and
\IEEEauthorblockN{Keke Zu}
\IEEEauthorblockA{Ericsson Research, Sweden\\
Email: zukeke@gmail.com  } \vspace{-1.75em}}

 \maketitle
\begin{abstract}
\boldmath
In this paper, we propose successive optimization non-linear
precoding strategies for physical-layer security in wireless
networks.  We also investigate different precoding techniques for
multi-user MIMO systems under various conditions of channel state
information (CSI) between the access point and the users and the
eavesdroppers. A non-linear precoding technique based on Successive
Optimization Tomlinson-Harashima Precoding (SO-THP) and Simplified
Generalized Matrix Inversion (S-GMI) technique is proposed along
with a strategy for injecting artificial noise prior to
transmission. Simulation results show that the proposed SO-THP+S-GMI
precoding technique outperforms existing non-linear and linear
precoding algorithms in terms of
BER and secrecy rate performances.\\

\end{abstract}
\section{Introduction}

Data security in wireless systems has been dominated in the past by
encryption methods like Data Encryption Standard (DES) and Advanced
Encryption Standard (AES) which have played very important roles in
the security aspects of data transmission. However, these encryption
algorithms suffer from high complexity and latency. This has
motivated the investigation of novel secrecy techniques which can be
implemented in the physical layer and which are under consideration
to provide secure wireless transmission.

In the 1970s, Wyner posed the Alice-Bob-Eve problem and proposed the
wire-tap transmission system in \cite{wyner}. The wire-tap
transmission system gives a basic description on how to achieve the
secrecy in the physical layer with better Gaussian channels between
users than the Gaussian wire-tap channel. Later on another study
reported in \cite{csiszar} showed that secrecy transmission is
possible even when the eavesdropper has a better channel to the
transmitter in a statistical sense than the receiver with
confidential messages. The secrecy capacity \cite{hellman} for
different kinds of channels, such as the Gaussian wire-tap channel,
the MIMO wire-tap channels have been studied in \cite{hellman} and
\cite{oggier}. In some later works \cite{goel}, \cite{mukherjee}, it
has been found that by adding artificial noise to the system the
secrecy of the transmission can be further enhanced.

\subsection{Related Work}

In recent years, precoding techniques have been widely applied to
multi-user MIMO systems. Precoding techniques rely on pre-processing
the transmitted data with the help of channel state information
(CSI). Different kinds of linear precoding techniques such as Zero-forcing (ZF) precoding, Minimum
Mean Square Error (MMSE) precoding and Block Diagonalization (BD)
precoding are introduced in \cite{stankovic}-\cite{geraci},
respectively. Some non-linear precoding techniques like
Tomlinson-Harashima precoding (THP) \cite{payaro}, Vector
Perturbation (VP) precoding \cite{sebdani} have also been reported.
The precoding techniques can also contribute to the performance of relay systems \cite{cai}
These precoding techniques have been proven effective in improving
the sum-rate and the bit-error rate (BER) performances. However,
precoding strategies require CSI at the transmit side, whereas in
reality  CSI is available to the users of interest but is rarely
available to the eavesdroppers. In most situations, a designer only
has access to imperfect CSI for the users of interest. Dealing with
imperfect CSI and the absence of CSI of the eavesdropper is a major
challenge. Moreover, a technique that is particularly effective for
improving the secrecy rate of physical-layer security is the
introduction of artificial noise at the transmitter \cite{goel}.

\subsection{Contributions}

In this pape we propose a novel non-linear technique based on
Successive Optimization Tomlinson-Harashima Precoding (SO-THP) and
Simplified Generalized Matrix Inversion (S-GMI) techniques along
with a strategy of injecting artificial noise prior to transmission.
We also study precoding techniques for multi-user MIMO systems under
various conditions of CSI between the access point and the users and
the eavesdroppers. Simulation results show that the proposed
SO-THP+S-GMI precoding technique outperforms existing non-linear and
linear precoding algorithms in terms of physical layer secrecy rate
performances.

The remainder of this paper is organized as follows. In Section II,
the system model and the performance metrics are given. The proposed
SO-THP+S-GMI precoding algorithm is introduced in Section III.
In Section IV, we discuss the simulation results.
The conclusions are given in Section V.

Notation: Bold uppercase letters ${\boldsymbol A}\in
{\mathbf{C}}^{M\times N}$ denote matrices with size ${M\times N}$
and bold lowercase letters ${\boldsymbol a}\in {\mathbf{C}}^{M\times
1}$ denote column vectors with length $M$. Conjugate, transpose, and
conjugate transpose are represented by $(\cdot)^\ast$, $(\cdot)^T$
and $(\cdot)^H$ respectively; $\boldsymbol I_{M}$ is the identity
matrix of size $M\times M$; $\rm diag \{\boldsymbol a\}$ denotes a
diagonal matrix with the elements of the vector $\boldsymbol a$
along its diagonal; $\mathcal{CN}(0,\sigma_{n}^{2})$ represents
$i.i.d$ entries with zero mean and $\sigma_{n}^{2}$ variance.

\section{System Model and Performance Metrics}

In this section we introduce the system model considered for the
transmission as well as the performance metrics that are used in the
evaluation of the precoders.

\subsection{System Model}
\begin{center}
\begin{figure}[h]
\vspace{-20pt}
\includegraphics[scale=0.55]{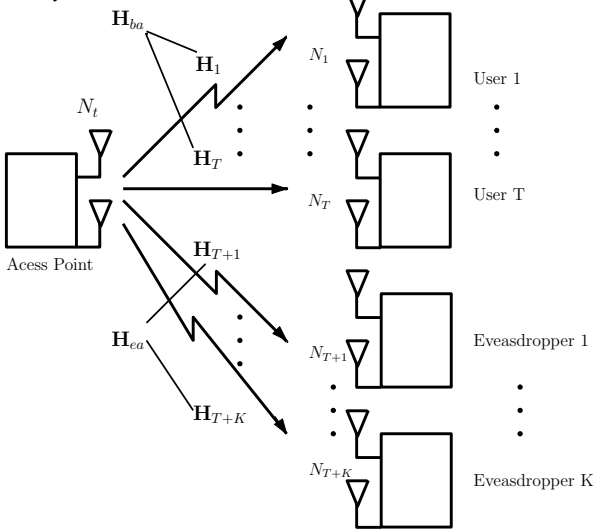}
\caption{System model of a MU-MIMO system with T users and K eavesdroppers}
\label{fig:sys}
\end{figure}
\vspace{-25pt}
\end{center}
Here we consider an uncoded multi-user multiple input multiple
output (MU-MIMO) broadcast channel. At the Acess Point (AP) a
transmitter with $N_{t}$ antennas is used to transmit $T$ users'
data steams. Each user's data streams are received with $N_{i}$
receive antennas. Along with the T users, $K$ eavesdroppers each
with $N_{k}$ receive antennas also receive data from the
transmitter. In this system we assume the eavesdroppers do not jam
the transmission and each user's or eavesdropper's channel is a
flat-fading MIMO channel. The quantity ${\boldsymbol H}_{i}\in
{\mathbf{C}}^{N_{i}\times N_{t}}$ and ${\boldsymbol H}_{k}\in
{\mathbf{C}}^{N_{k}\times N_{t}}$ corresponds to each user's or the
eavesdropper's channel matrix. The Gaussian noise with $i.i.d$
entries is distributed as $\mathcal{CN}(0,\sigma_{B}^{2})$ and
$\mathcal{CN}(0,\sigma_{E}^{2})$.

In the scenarios without artificial noise, each user's or
eavesdropper's received data are described by
\begin{equation}
{\boldsymbol y}_{r}=\beta_{r}^{-1}({\boldsymbol H}_{r}{\boldsymbol P}_{r} {\boldsymbol s}_{r} + {\boldsymbol H}_{r}\sum \limits_{j=1,j\neq r}^R
{\boldsymbol P}_{j} {\boldsymbol s}_{j}+ {\boldsymbol n}_{r}),\label{eqn:yr}
\end{equation} in (\ref{eqn:yr})
$\beta_{r}=\sqrt{\dfrac{E_{r}}{||\boldsymbol P_{r}||}}$ is used to
guarantee that the transmit power after precoding is the same as the
original transmit power $E_{r}$ of user $r$
 and ${\boldsymbol P}_{r}\in {\mathbf{C}}^{N_{t}\times N_{r}}$ is
the precoding matrix. We use the vector ${\boldsymbol x}_{r}\in
{\mathbf{C}}^{N_{t} \times 1}$ to represent the signal after
precoding. Without artificial noise ${\boldsymbol x}_{r}$ is given
by
\begin{equation}
{\boldsymbol x}_{r}={\boldsymbol P}_{r} {\boldsymbol s}_{r},
\label{eqn:xr1}
\end{equation}
where ${\boldsymbol s}_{r}\in
{\mathbf{C}}^{N_{r}\times 1}$ is the transmit symbols.

If the artificial noise is applied before the transmission,
following the description in \cite{mukherjee}, the signal after
precoding is given by
\begin{equation}
{\boldsymbol x}_{r}={\boldsymbol P}_{r}{\boldsymbol s}_{r}+
{\boldsymbol P}_{r}' {\boldsymbol s}_{r}',
\label{eqn:xr2}
\end{equation}
where ${\boldsymbol P}_{r}\in {\mathbf{C}}^{N_{t}\times N_{r}}$,
${\boldsymbol P}_{r}'\in {\mathbf{C}}^{N_{t}\times m}$ are the
precoding matrices. The vector ${\boldsymbol s}_{r}\in
{\mathbf{C}}^{N_{r}\times 1}$ represents the information to be
transmitted. The vector ${\boldsymbol s}_{r}'\in
{\mathbf{C}}^{m\times 1}$ is the jamming signal. Suppose $0<\rho<1$
is the fraction of the power devoted to the information signal and
${{\boldsymbol P}_{r}'}^H {{\boldsymbol P}_{r}'}=\boldsymbol I$. The
artificial noise then should satisfy
\begin{equation}
E\{{{\boldsymbol s}_{r}'} {{\boldsymbol s}_{r}'}^{H}\}=\boldsymbol Q_{s}',
\label{eqn:eqs}
\end{equation}
\begin{equation}
Tr({\boldsymbol Q}_{s}')=(1-\rho)E_{s},
\label{eqn:trqs}
\end{equation}
In (\ref{eqn:eqs}) and (\ref{eqn:trqs}) ${\boldsymbol Q}_{s}' \in
{\mathbf{C}}^{m\times m}$ is the covariance matrix associated with
the jamming signal $s_{r}'$. The covariance matrix of the transmit
signal ${\boldsymbol Q}_{s} \in {\mathbf{C}}^{N_{r}\times N_{r}}$
satisfies
\begin{equation}
Tr({\boldsymbol Q}_{s})=\rho E_{s},
\end{equation}
and proper artificial noise power ratio $\rho$  will result in an
increase of the secrecy rate. Based on the construction of the
artificial noise, there will be a different effect on the secrecy
rate performance. In \cite{lin}, it shows a generalized  artificial
noise with more flexible covariance matrix to improve the secrecy
rate. Besides the construction introduced in \cite{mukherjee}, the
study in \cite{chorti} reports a system with a helping interferer to
improve the achievable secrecy rate.

\subsection{Performance Metrics}

From \cite{wyner}-\cite{oggier}, we have the definition of the
secrecy capacity as the maximum transmission rate Alice can achieve
during the transmission without any information detected by the
eavesdropper. From \cite{oggier} without artificial noise, we have
the secrecy capacity
\begin{equation}
\begin{split}
C_{sec}&=\max_{{\boldsymbol Q}_{s}\geq 0 , \rm Tr({\boldsymbol Q}_{s})\leq E_{s}}{\textit{I}(X_{s};Y_{i})-\textit{I}(X_{s};Y_{k})}\\
             &=\max_{{\boldsymbol Q}_{s}\geq 0, \rm Tr({\boldsymbol Q}_{s})\leq E_{s}}\log(\det({\boldsymbol I}+{\boldsymbol H}_{ba} {\boldsymbol Q}_{s} {\boldsymbol H}_{ba}^H/\sigma_{B}^{2}))\\
             & \quad -\log(\det({\boldsymbol I}+ {\boldsymbol H}_{ea}{\boldsymbol Q}_{s} {\boldsymbol H}_{ea}^H/\sigma_{E}^{2})),
\end{split}
\label{eqn:Cs}
\end{equation}
In (\ref{eqn:Cs}) $\textit{I}(X_{s};Y_{i})$ and
$\textit{I}(X_{s};Y_{k})$ represent the mutual information between
the transmitter to the user as well as the eavesdropper.
$X_{s},Y_{i},Y_{k}$ are the counterparts of different signal
symbols. ${\boldsymbol Q}_{s}$ is the covariance matrix associated
with the signal after precoding ${\boldsymbol x}_{r}$. The secrecy
capacity achieved in this situation with prior knowledge of the
users' channel ${\boldsymbol H}_{ba}$ as well as the eavesdroppers'
channel ${\boldsymbol H}_{ea}$ can be used as a benchmark for
further study. In reality the channel between the transmitter and
the eavesdroppers is not perfectly known. This situation known as
the imperfect CSI is discussed in \cite{goel}. However, for the
imperfect CSI  ${\boldsymbol H}_{ea}$ we suppose the distribution of
the eavesdroppers' channel is known to the transmitter and users.

Another performance metric is the BER estimated by the different
links. Ideally, we would like the users to experience a reliable
communication and the eavesdropper to have a very high BER
(virtually no reliability when communicating). In this paper,
different precoding techniques are supposed to suppress the
interferences between different users thus leading to a better BER
performance.

Thirdly, the algorithm computational complexity should also take
into consideration. With the increasing of the number of users, the
high computational complexity algorithm will suffer from a long time
delay. Low computational complexity will contribute to reduce the
latency of the system.

\section{Proposed  Precoding Algorithms}

In this section, we detail the proposed precoding algorithms.

\subsection{SO-THP Precoding}

According to \cite{shenouda}, we assume the modified date symbol vector $\boldsymbol u_{r}=\boldsymbol s_{r}+\boldsymbol z_{r}$, where $\boldsymbol u_{r} \in {\mathbf{C}}^{N_{r}\times 1}$ and $\boldsymbol z_{r} \in {\mathbf{C}}^{N_{r}\times 1}$.
With the feedback matrix we can then obtain the output of the modulo operation $\boldsymbol v_{r}$
\begin{equation}
\boldsymbol v_{r}=(\boldsymbol I + \boldsymbol B)_{r}^{-1} \boldsymbol u_{r}
\label{eqn:vr}
\end{equation}
where $\boldsymbol v_{r} \in {\mathbf{C}}^{N_{r}\times 1}$. When we have the feedforward matrix $\boldsymbol F_{r} \in {\mathbf{C}}^{N_{t}\times N_{r}}$,
the transmit signal can be obtained as $\boldsymbol x_{r}=\boldsymbol F_{r} \boldsymbol v_{r}$. where $\boldsymbol x_{r} \in {\mathbf{C}}^{N_{t}\times 1}$
The received signal $\boldsymbol y_{r-pro} \in {\mathbf{C}}^{N_{t}\times 1}$ then can be written in the following form:
\begin{equation}
\boldsymbol y_{r-pro}=\boldsymbol H_{r} \boldsymbol x_{r}+\boldsymbol n_{r}\\
=\boldsymbol H_{r} \boldsymbol F_{r}(\boldsymbol I + \boldsymbol B)_{r}^{-1} \boldsymbol u_{r}+\boldsymbol n_{r}
\label{eqn:yrpro}
\end{equation}
In \cite{haardt}, the SO-THP algorithm is illustrated as a
combination of a successive optimization technique and
Tomlinson-Harashima precoding. The reordered feedforward and
feedback matrices are expressed as
\begin{equation}
{\boldsymbol F}= \left( {\boldsymbol F}_{1}  \cdots  {\boldsymbol F}_{R}  \right)
\label{eqn:fmatrix}
\end{equation}
\begin{equation}
{\boldsymbol B}={\rm lower~triangular}\left({\boldsymbol D}{\boldsymbol H}{\boldsymbol F}\bullet {\rm diag}\left([{\boldsymbol D}{\boldsymbol H}{\boldsymbol F}]_{ii}^{-1}\right)\right)
\label{eqn:bmatrix}
\end{equation}
where the complex unitary matrix ${\boldsymbol D}=\begin{pmatrix}
 {\boldsymbol D}_{1} &         & &\\
         & \ddots & &  \\
         &         & {\boldsymbol D}_{T} &\\ \end{pmatrix} $.

\subsection{S-GMI Precoding}

The concept of generalized matrix inversion precoding has been
introduced in \cite{hakjea}, where two approaches called Generalized
Zero-Forcing Channel Inversion (GZI) and Generalized MMSE Channel
Inversion (GMI) are illustrated. The GMI scheme uses the QR
decomposition to decompose the MMSE channel inversion
$\bar{{\boldsymbol H}_{r}}$ as expressed by
\begin{equation}
\bar{{\boldsymbol H}_{r}}=({{\boldsymbol H}_{r}^H}{\boldsymbol H}_{r}+\alpha \boldsymbol I)^{-1}{\boldsymbol H}_{r}^H,
\label{eqn:parh}
\end{equation}
\begin{equation}
\bar{{\boldsymbol H}_{r}}=\bar{{\boldsymbol Q}_{r}}\bar{{\boldsymbol R}_{r}} \qquad  for \qquad r=1,\cdots,R,
\label{eqn:parhqr}
\end{equation}
where $\bar {{\boldsymbol H}_{r}}\in {\mathbf{C}}^{N_{t}\times
N_{r}}$, $\bar{{\boldsymbol Q}_{r}}\in {\mathbf{C}}^{N_{t}\times
N_{t}}$, $\bar{{\boldsymbol R}_{r}}\in {\mathbf{C}}^{N_{t}\times
N_{r}}$
and GMI takes the noise into account. It overcomes the noise
enhancement in the BD algorithm. Meanwhile extra interference may be
introduced in the following steps. To solve this problem, a transmit
combining matrix ${\boldsymbol T}_{r}$ is applied to
$\bar{{\boldsymbol Q}_{r}}$. Under the total transmit power
constraint we use the minimum total MSE criterion to achieve the
transmit combining matrix ${\boldsymbol T}_{r}$. In \cite{hakjea}
the transmit combining matrix ${\boldsymbol T}_{r}$ is given by
${\boldsymbol T}_{r}=\beta \bar{{\boldsymbol T}_{r}}$ as described
by
\begin{equation}
\bar{{\boldsymbol T}_{r}}=\left(\bar{{\boldsymbol Q}_{r}}^H \sum\limits_{j=1}^R {\boldsymbol H}_{j}^H {\boldsymbol H}_{j}\bar{{\boldsymbol Q}_{r}}+\alpha \boldsymbol I  \right)^{-1} \bar{{\boldsymbol Q}_{r}}^H {\boldsymbol H}_{r}^H {\boldsymbol H}_{r}\bar{{\boldsymbol Q}_{r}},
\label{eqn:Tr}
\end{equation}
where ${{\boldsymbol T}_{r}}$, $ \bar{{\boldsymbol T}_{r}}\in {\mathbf{C}}^{N_{t}\times N_{t}}$.
Once we have $\bar{{\boldsymbol Q}_{r}}$ and ${\boldsymbol T}_{r}$, the following approach is
\begin{equation}
{{\boldsymbol H}_{r}}\bar{{\boldsymbol Q}_{r}}{\boldsymbol T}_{r}={\bar{{\boldsymbol U}_{r}}}{\bar{{\boldsymbol \Sigma}_{r}}}{\bar{{\boldsymbol V}_{r}}}^H,
\label{eqn:Trsvd}
\end{equation}
In \cite{keke1}, it has been shown that the transmit combining
matrix $\boldsymbol T_{r}$ is not necessary. Therefore, a simplified
GMI (S-GMI) is developed in \cite{keke1} as an improvement of the
original RBD precoding in \cite{stankovic}. This is known as S-GMI.
Therefore, the decomposition  in (\ref{eqn:Trsvd}), the precoder and
the receive filter are
\begin{equation}
{{\boldsymbol H}_{r}}\bar{{\boldsymbol Q}_{r}}=\tilde{{\boldsymbol U}_{r}}\tilde{{\boldsymbol \Sigma}_{r}}\tilde{{{\boldsymbol V}_{r}}}^H,
\label{eqn:sgmisvd}
\end{equation}
\begin{equation}
{{\boldsymbol P}_{S-GMI}}=[\bar{{\boldsymbol Q}_{1}}\tilde{{\boldsymbol V}_{1}} \quad \bar{{\boldsymbol Q}_{2}}\tilde{{\boldsymbol V}_{2}}\quad \cdots \quad \bar{{\boldsymbol Q}_{R}}\tilde{{\boldsymbol V}_{R}}],
\label{eqn:psgmi}
\end{equation}
\begin{equation}
{{\boldsymbol M}_{S-GMI}}={\rm diag}\{\tilde{{\boldsymbol U}_{1}}^H \quad \tilde{{\boldsymbol U}_{2}}^H \quad \cdots \quad \tilde{{\boldsymbol U}_{R}}^H\},
\label{eqn:msgmi}
\end{equation}
where ${{\boldsymbol P}_{S-GMI}}\in {\mathbf{C}}^{N_{t}\times
N_{t}}$, ${{\boldsymbol M}_{S-GMI}}\in {\mathbf{C}}^{N_{t}\times
N_{t}}$. In the SO-THP+S-GMI algorithm, the feedforward matrix
(\ref{eqn:fmatrix}) is obtained by reordering the S-GMI precoding
matrix (\ref{eqn:psgmi}).

\section{Simulation Results}

A system with $N_{t}=4$ transmit antennas and $T=2$ users as well as
$K=2$ eavesdroppers is considered. All the transmitted symbols are
modulated with QPSK. In the simulations, we assume that the channel
for each user is uncorrelated and is generated following a complex
Gaussian model with zero mean and variance equal to one.

\subsection{Computational complexity}

In assessing the computational complexity, the number of  FLOPS
(Floating-Point Operations Per Second) is used as a measurement of
the cost of an algorithm. In \cite{golub}, the FLOPS for real QR,
SVD and complex QR algorithms are given. According to \cite{keke2}
for an $m\times n$ complex matrix $\boldsymbol A$ ,the FLOPS of the
SVD of the complex matrix $\boldsymbol A$ are equivalent to its
$2m\times 2n$ extended real matrix. The number of FLOPS of the
multiplication, the QR decomposition, the matrix inversion as well
as SVD with different matrices obtained have been reported in
\cite{golub}.

\begin{center}
\begin{figure}[h]
\vspace{-5pt}
\includegraphics[scale=0.5]{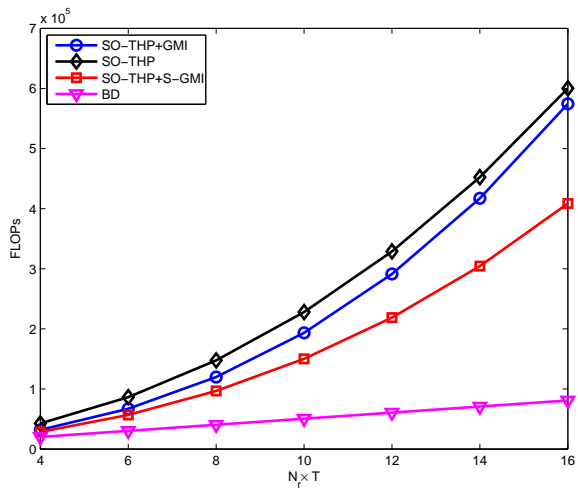}
\caption{Computational complexity in FLOPs for MU-MIMO systems}
\label{fig:com}
\end{figure}
\vspace{-25pt}
\end{center}

In Fig. \ref{fig:com} it is shown that although the BD precoding
algorithm has the lowest complexity, its BER and secrecy rate
performances are not satisfactory. Among the SO-THP type algorithms
the SO-THP+S-GMI algorithm shows an advantage over the standard
SO-THP algorithm in terms of complexity.

\begin{center}
\begin{figure}[h]
\vspace{-5pt}
\includegraphics[scale=0.5]{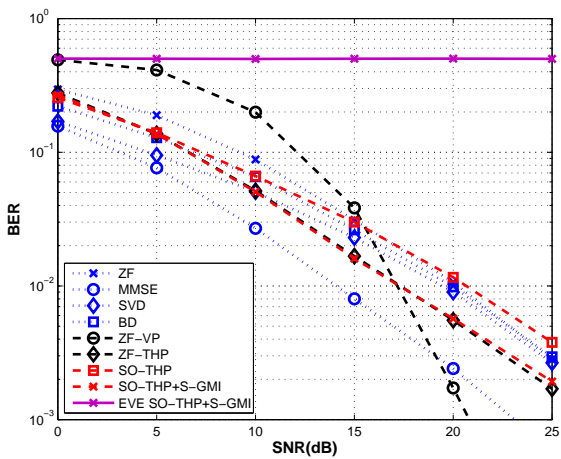}
\caption{BER performance with precoding techniques in $4 \times 4 \times 4$ MU-MIMO broadcast channel, $m=0.5$}
\label{fig:ber1}
\end{figure}
\vspace{-20pt}
\end{center}

\subsection{Perfect Channel State Information}

Fig. \ref{fig:ber1} shows the BER performance for different
precoding techniques. The proposed SO-THP+S-GMI precoding requires
less than 5 dB than the original SO-THP precoding in terms of SNR to
achieve $10^{-2}$ bit error rate. According to Fig. \ref{fig:444P},
the secrecy rate will level out according to the ratio between the
main channel and the wire-tap channel $m$. From Fig. \ref{fig:444P},
we can also see that among all the investigated precoding
algorithms, the proposed SO-THP+S-GMI precoding can achieve a
certain secrecy rate with lower SNR.

\begin{center}
\begin{figure}[h]
\vspace{-5pt}
\includegraphics[scale=0.5]{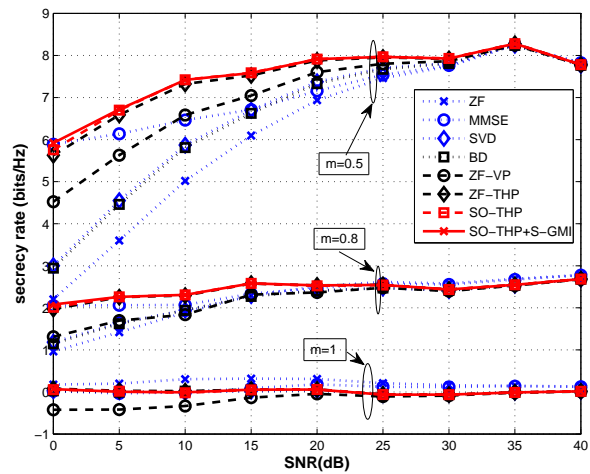}
\caption{Secrecy rate performance with precoding techniques in $4 \times 4 \times 4$ MU-MIMO broadcast channel}
\label{fig:444P}
\end{figure}
\vspace{-20pt}
\end{center}

\begin{center}
\begin{figure}[h]
\vspace{-5pt}
\includegraphics[scale=0.5]{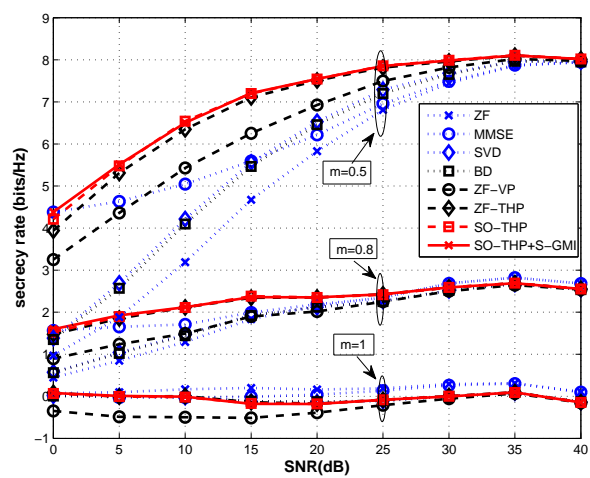}
\caption{Secrecy rate with precoding techniques $4 \times 4 \times 4$ MU-MIMO broadcast channel with imperfect CSI}
\label{fig:444IP}
\end{figure}
\end{center}

\subsection{Imperfect Channel State Information}

In Fig. \ref{fig:444IP}, the secrecy rate performance is evaluated
in the imperfect CSI scenario, which shows a degradation at low SNR.
Furthermore, the secrecy rate requires a very high SNR to converge
to a high secrecy rate. The proposed SO-THP+S-GMI precoder has the
best secrecy rate performance among all the studied precoding
techniques.

\begin{center}
\begin{figure}[h]
\vspace{-5pt}
\includegraphics[scale=0.5]{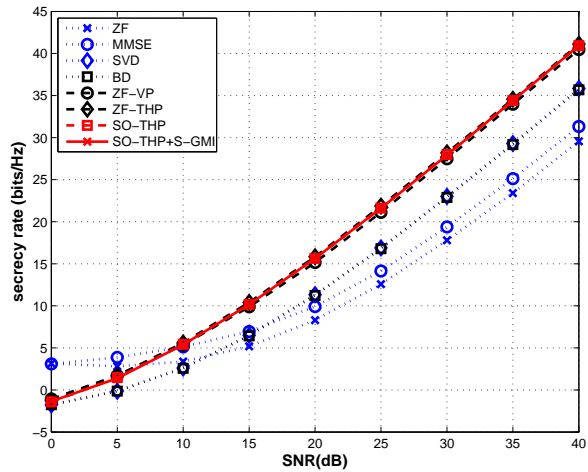}
\caption{Secrecy rate with precoding techniques $4 \times 4 \times
4$ MU-MIMO broadcast channel with imperfect CSI adding artificial
noise, $m=0.5$} \label{fig:444AIP}
\end{figure}
\vspace{-20pt}
\end{center}

\begin{center}
\begin{figure}[h]
\vspace{-5pt}
\includegraphics[scale=0.5]{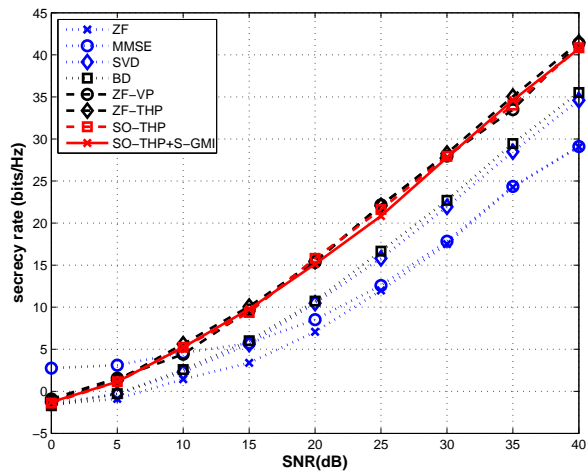}
\caption{Secrecy rate with precoding techniques $4 \times 4 \times 4$ MU-MIMO broadcast channel with imperfect CSI adding artificial noise, $m=2$}
\label{fig:444m2AIP}
\end{figure}
\vspace{-20pt}
\end{center}

\subsection{Imperfect Channel State Information With Artificial Noise}

In Fig. \ref{fig:444AIP} and Fig. \ref{fig:444m2AIP}, we assume that
artificial noise is generated with extra free antennas. Although the
artificial noise is added, the total transmit power $E_{s}$ will
remain the same. In the simulation $0.4$ of the transmit power
$E_{s}$ is used to generate the artificial noise. In Figure
\ref{fig:444AIP}, with higher SNR, we can have better secrecy rate
performance. Even when the eavesdropper has better statistical
channel, the secrecy rate keeps almost the same according to Figure
\ref{fig:444m2AIP}.

\section{Conclusion}

Precoding techniques used in MU-MIMO broadcast systems can achieve
good BER and secrecy-rate performances. The proposed SO-THP+S-GMI
algorithm requires the lowest computational complexity among
non-linear techniques and outperforms most linear and non-linear
techniques in BER performance. Moreover, SO-THP+S-GMI can achieve a
high secrecy rate, which means the security of the transmission with
SO-THP+S-GMI is superior to existing techniques. When artificial
noise is added, the performance is further improved.

\end{document}